\documentclass[aps,pra,twocolumn,nobibnotes]{revtex4-1}
\usepackage{graphicx}
\usepackage{multirow,bigdelim}
\usepackage{bm}
\usepackage{amsmath}
\usepackage{lipsum}
\usepackage[breaklinks=true]{hyperref}

\hypersetup{colorlinks=true,
pdftitle={Effect of line broadening on the performance of atomic Faraday filters},
pdfauthor={Mark Zentile},
linkcolor=blue,
citecolor=blue,
urlcolor=blue}

\newcommand{\mrm}{\mathrm}

\begin{document}

\title{Effect of line broadening on the performance of Faraday filters}
\date{\today}
\author{Mark A Zentile}
\email{m.a.zentile@durham.ac.uk}
\affiliation{Joint Quantum Centre (JQC) Durham-Newcastle, Department of Physics, Durham University, South Road, Durham, DH1 3LE, United Kingdom}
\author{Renju S Mathew}
\affiliation{Joint Quantum Centre (JQC) Durham-Newcastle, Department of Physics, Durham University, South Road, Durham, DH1 3LE, United Kingdom}
\author{Daniel J Whiting}
\affiliation{Joint Quantum Centre (JQC) Durham-Newcastle, Department of Physics, Durham University, South Road, Durham, DH1 3LE, United Kingdom}
\author{James Keaveney}
\affiliation{Joint Quantum Centre (JQC) Durham-Newcastle, Department of Physics, Durham University, South Road, Durham, DH1 3LE, United Kingdom}
\author{Charles S Adams}
\affiliation{Joint Quantum Centre (JQC) Durham-Newcastle, Department of Physics, Durham University, South Road, Durham, DH1 3LE, United Kingdom}
\author{Ifan G Hughes}
\affiliation{Joint Quantum Centre (JQC) Durham-Newcastle, Department of Physics, Durham University, South Road, Durham, DH1 3LE, United Kingdom}

\begin{abstract}
We show that homogeneous line broadening drastically affects the performance of atomic Faraday filters. We use a computerized optimization algorithm to find the best magnetic field and temperature for Faraday filters with a range of cell lengths. The effect of self-broadening is found to be particularly important for short vapour cells, and for `wing-type' filters. Experimentally we realize a Faraday filter using a micro-fabricated $^{87}$Rb vapour cell. By modelling the filter spectrum using the ElecSus program we show that additional homogeneous line broadening due to the background buffer-gas pressure must also be included for an accurate fit.
\end{abstract}

\maketitle

\section{Introduction}

Devices utilising thermal atomic vapour cells are of increasing interest since they offer high precision with a compact and relatively simple apparatus. Examples of atomic vapour cell devices include magnetometers~\cite{Kominis2003,Budker2007}, gyroscopes~\cite{Lam1983,Kornack2005}, clocks~\cite{Knappe2004a,Camparo2007}, electric field sensors~\cite{Mohapatra2008}, microwave detectors~\cite{Sedlacek2012,Sedlacek2013} and cameras~\cite{Bohi2012,Horsley2013a,Fan2014a}, quantum memories~\cite{Julsgaard2004,Lvovsky2009,Sprague2014}, optical isolators~\cite{Weller2012d}, laser frequency references~\cite{Affolderbach2005} and narrowband optical notch~\cite{Miles2001,Uhland2015} and bandpass filters~\cite{Ohman1956,Beckers1970}.

Making these devices more compact, power efficient and lighter is currently a burgeoning area of research~\cite{Mescher2005,Ompact2008,Mhaskar2012}, since it allows them to become practical consumer products. Particularly for devices that require an applied magnetic field, compact vapour cells~\cite{Sarkisyan2001,Liew2004,Knappe2005,Su2009,Baluktsian2010,Tsujimoto2013,Straessle2014} offer the additional advantage that small permanent magnets can be used to create a uniform magnetic field across the vapour cell~\cite{Weller2012c}, while consuming no power. However, when confining the atomic vapour in small geometries, additional effects may need to be taken into account. For example, atom-surface interactions become important for atoms in hollow-core fibres~\cite{Epple2014} or nano-metric thin cells~\cite{Whittaker2014}. Also, cells with a shorter path length require the medium to be heated more to increase the atomic number density. Not only will this increased heating cause more Doppler broadening but the increased number density will mean that self-broadening~\cite{Lewis1980,Weller2011} must be taken into account. In this article we investigate the effects of these homogeneous and inhomogeneous broadening mechanisms on the performance of Faraday filters.

Faraday filters were proposed in 1956 by \"{O}hman~\cite{Ohman1956} for astrophysical observations. They were later applied to solar observations~\cite{Agnelli1975,Cacciani1978} and used to frequency stabilize dye lasers~\cite{Sorokin1969,Yabuzaki1977,Endo1978}. In the early 1990s the subject of Faraday filters was revived~\cite{Dick1991,Menders1991}. Such filters have received increasing attention ever since, owing to their high performance in many applications. Faraday filters now find use in remote temperature sensing~\cite{Popescu2004}, atmospheric lidar~\cite{Chen1996,Fricke-Begemann2002,Huang2009,Harrell2010}, diode laser frequency stabilisation~\cite{Wanninger1992,Choi1993,Miao2011}, Doppler velocimetry~\cite{Cacciani1978,Bloom1991,Bloom1993}, communications~\cite{Junxiong1995} and quantum key distribution~\cite{Shan2006} in free space, optical limitation~\cite{Frey2000}, filtering Raman light~\cite{Abel2009}, and quantum optics experiments~\cite{Siyushev2014,Zielinska2014a}.

The Faraday-filter spectrum is sensitive to many experimental parameters and so a theoretical model is useful for designing filters. However, there are only a few articles describing computer optimization~\cite{Kiefer2014,Zentile}. In this article we use computer optimization to find the best working conditions for compact Faraday filters. We find homogeneous broadening is particularly important for Faraday filters in `wing' operation~\cite{Zielinska2012,Zentile} and less so for `line-centre' operation~\cite{Chen1993,Kiefer2014}. The homogeneous broadening mechanism of self-broadening is particularly important to include since it is unavoidable at high density. Previous theoretical treatments of Faraday filters~\cite{Yin1991,Harrell2009,Zielinska2012} have not included the effect of self-broadening; we find that self-broadening is important for short cell lengths and must be included in the model in order to find the best working parameters.   The structure of the rest of the article is as follows: In section~\ref{sec:Theory} we introduce the typical experimental arrangement for Faraday filters and qualitatively explain how they work. In section~\ref{sec:Opt} we explain the computer optimization technique used to find the best working parameters and show the importance of self-broadening for shorter cells. Section~\ref{sec:Exp} describes an experiment performed to compare with the theoretical optimizations. The results show that buffer gas broadening and isotopic purity strongly effect the filter spectrum. Finally we draw our conclusions in section~\ref{sec:Conc}.

\section{Theory and Background}\label{sec:Theory}

\begin{figure}[t]
\includegraphics[width=\columnwidth]{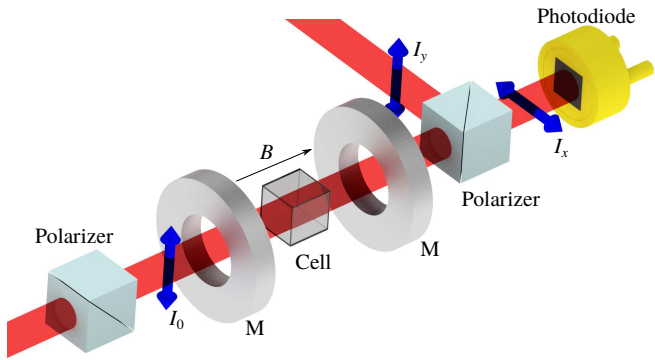}
\caption{Illustration of the experimental arrangement. A micro-fabricated $1\times1\times1\,$mm$^3$ $^{87}$Rb vapour cell is placed between two axially magnetized ring magnets. This arrangement is then placed between two crossed polarizers, forming the filter. The filter is tested by passing a laser beam through and onto a photodiode. The filter transmission is defined as the intensity of light transmitted through the second polarizer ($I_x$) divided by the initial intensity before the cell ($I_0$). Light out of the passband frequency is either scattered in the cell or rejected at the second polarizer ($I_y$).}
\label{fig:setup}
\end{figure}

An atomic Faraday filter is formed by surrounding an atomic vapour cell with crossed polarizers (see figure~\ref{fig:setup}). When an axial magnetic field ($B$) is applied across the cell, the medium becomes circularly birefringent causing the plane of polarization to rotate as light traverses the cell (the Faraday effect~\cite{Budker2002}), which leads to some transmission through the second polarizer. For a dilute atomic medium the effect is negligibly small except near resonances, and since atomic resonances are extremely narrow, this results in a narrowband filter. If the signal being detected is unpolarized then half of the light will not pass through the first polarizer. This limits the filter transmission to 50\%. However using a polarizing beam splitter allows one to arrange two Faraday filters to allow each polarization component through with little loss~\cite{Fricke-Begemann2002}.

In a similar way, if the magnetic field is perpendicular to the light propagation direction, one can also make a `Voigt filter'~\cite{Menders1992} which exploits the Voigt effect~\cite{Franke-Arnold2001}. However, in this paper we will only consider Faraday filters. We have chosen to consider the D$_2$ ($\mrm{n}^2\mrm{S}_{1/2}\rightarrow \mrm{n}^2\mrm{P}_{3/2}$) lines of potassium and rubidium where $\mrm{n}=$ 4 or 5 respectively.

For a given cell length the parameters that affect the Faraday filter transmission spectra are the applied field ($B$) and cell temperature ($T$). The effect of $T$ is predominantly to change the atomic number density~\cite{Alcock1984} and secondly Doppler width, while $B$ causes the circular birefringence and dichroism. In general the filter spectrum is a complicated function of these two parameters, due to the large number of non-degenerate Zeeman shifted transitions, each with different transition strengths in which their lineshape profiles partially overlap. However, it is possible to accurately compute the filter profile with a computer program~\cite{Zentile2014a,Zielinska2012,Kiefer2014}. 

We use the ElecSus program to calculate the filter spectrum. The full description of how the program works can be found in ref.~\cite{Zentile2014a}; here we summarize the key points. An atomic Hamiltonian is built up from contributions from hyperfine and magnetic interactions. The eigenvalues allow the transition frequencies to be calculated while the eigenstates can be used to calculate their strengths. The electric susceptibility is then calculated by adding the appropriate (complex) line-shape at each transition frequency, scaled by its strength. The imaginary part of these line-shapes have a Voigt profile~\cite{Corney1977}, which is a convolution between inhomogeneous broadening (Gaussian profile from Doppler broadening) and homogeneous broadening (Lorentzian profile). Typically, the full-width half maximum of the Lorentzian has contributions from natural broadening ($\Gamma_0$) and self-broadening ($\Gamma_\mrm{self}$) and buffer gas pressures ($\Gamma_\mrm{buf}$). The real part of the electric susceptibility can be used to calculate dispersion, whilst the imaginary part can be used to calculate extinction~\cite{Jackson1999}. This allows the calculation of a variety of experimental spectra, of which the Faraday filter spectrum is one. The result is given as a function of global detuning, $\Delta$, which is defined as $\Delta \equiv \omega - \omega_0$, were $\omega$ is the angular frequency of the laser light and $\omega_0$ is the global line-centre angular frequency.

\section{Optimization}\label{sec:Opt}

\subsection{The simple approach}\label{sec:Simple}

The optical signal in a vapour cell device comes from the interaction of the light with all the atoms in the beam path. This means that for compact vapour cells with shorter path lengths, the atomic number density must increase to compensate for the loss of signal. For example the Faraday filter spectrum can be thought of as some function of the product $\sigma\mathcal{N}L$, where $\mathcal{N}$ is the number density, $L$ is the length of the medium and $\sigma$ is the microscopic atomic cross-section (describing the effect of extinction and dispersion due to a single atom). Assuming $\sigma$ remains constant, we can achieve the same filter when reducing $L$ by increasing $\mathcal{N}$ by the same factor. Therefore, once good parameters of $B$ and $T$ are found for a particular cell length, we can find the new appropriate parameters by changing the temperature such that $\mathcal{N}L$ remains constant.

However, this argument will break down at some point since $\sigma$ is not generally constant. By increasing the cell temperature we also change the amount of Doppler broadening. Also, at high densities, interactions between atoms cause self-broadening, which can be modelled as $\Gamma_\mrm{self}=\beta\mathcal{N}$, where $\beta$ is the self-broadening parameter~\cite{Weller2011}. Both the Doppler and self-broadening will affect $\sigma$. To find where these effects become important we need to compare it with a computer optimization technique, which can find the best parameters at each cell length.

\subsection{Computerized optimization procedure}\label{sec:CompOpt}

Efficiently finding the optimal experimental conditions for a Faraday filter requires three tools. First a computer program is needed which can calculate the spectrum with the experimental conditions as parameters. Secondly, a definition of a figure of merit (or conversely a `cost function'~\cite{Russel2003}) is then needed to numerically quantify which filter spectra are more desirable. Finally, this figure of merit is then maximised (or the cost function is minimised) by varying the parameters according to some algorithm.

We used a global minimization technique~\cite{Hughes2010} which includes the random-restart hill climbing meta-algorithm~\cite{Russel2003} in conjunction with the downhill simplex method~\cite{Nelder1965} to find the values of $B$ and $T$ which maximized our figures of merit. This routine was used in conjunction with the ElecSus program~\cite{Zentile2014a} which calculated the filter spectra. ElecSus was used because it includes the effect of self-broadening, which is essential for this study, and also because it evaluates the filter spectrum quickly ($<1\,$s) which makes this kind of optimization practical, since the filter spectra need to be evaluated a few thousand times.

\subsection{Figure-of-merit choices}\label{sec:FOMs}

The signal-to-noise ratio of a narrowband signal in broadband noise is greatly improved by using a bandpass filter. For the case of white noise, the noise power is directly proportional to the bandwidth of a top-hat filter. For a more general filter profile, the equivalent-noise bandwidth (ENBW) is a quantity which is inversely proportional to the signal to noise ratio, and is defined as
\begin{equation}
\mrm{ENBW}=\frac{\int^\infty_0 I_x(\nu)\mrm{d}\nu}{I_x(\nu_s)},
\label{eq:ENBW}
\end{equation} 
where $I_x$ is the light intensity after the filter, $\nu$ is the optical frequency and $\nu_s$ is the signal frequency. If there is freedom in the exact position of the signal frequency we can set it to the frequency which gives the maximum transmission ($I_x(\nu_s)=I_\mrm{max}$).

\begin{figure}
\includegraphics[width=\columnwidth]{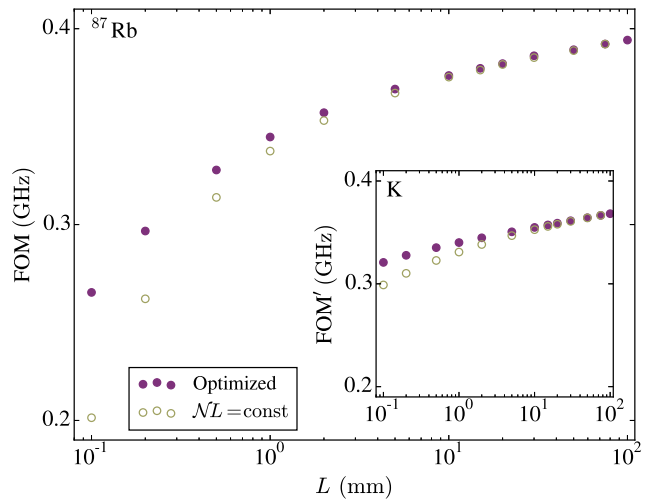}
\caption{The figures of merit of filter spectra found by optimization or extrapolation. The hollow (olive) circles show the figure of merit found by taking the optimal magnetic field and temperature of the 100 mm length cell and changing the temperature such that $\mathcal{N}L=\mrm{const}$. The solid (purple) dots show the figure of merit maximized by changing the magnetic field and temperature for each cell length. The main panel shows the results of a wing-type filter using an isotopically pure $^{87}$Rb vapour, the inset shows a line-centre filter with a potassium vapour at natural abundance. Both are modelled for the D$_2$ line of the respective element.}
\label{fig:FomOpt}
\end{figure}

Although minimising the ENBW is desirable, this usually comes with a reduction in transmission~\cite{Kiefer2014}. Using the following figure of merit,
\begin{equation}
\mrm{FOM} = \left.\frac{I_\mrm{max}^2}{\int^\infty_0 I_x(\nu)\mrm{d}\nu} = \frac{I_\mrm{max}}{\mrm{ENBW}}\right\rvert_{I_x(\nu_s)=I_\mrm{max}},
\label{eq:FOM1}
\end{equation}
we can maintain a reasonably large transmission~\cite{Kiefer2014}, while minimizing the ENBW. When optimising using this figure of merit we often find a wing-type filter spectrum~\cite{Zentile}. In order to compare with line-centre filters we also use the following figure of merit,
\begin{equation}
\mrm{FOM^\prime} = \left.\frac{I_x^2(\nu_s)}{\int^\infty_0 I_x(\nu)\mrm{d}\nu}\right\rvert_{\nu_s=\omega_0/2\pi},
\label{eq:FOM2}
\end{equation} 
where we set $\nu_s$ to be the line-centre frequency.

To calculate these figure-of-merit values we simulate filter spectra with a range of 60 GHz around the atomic weighted line-centre with a 10 MHz grid spacing. The integration is performed by a simple rectangle method. The limitation to the accuracy of calculated the figure-of-merit values comes from the grid spacing; a finer grid spacing of 1 MHz only improves the accuracy by 0.2\% at best.

\subsection{Results for wing and line-centre filters}

\begin{figure}%
\includegraphics[width=\columnwidth]{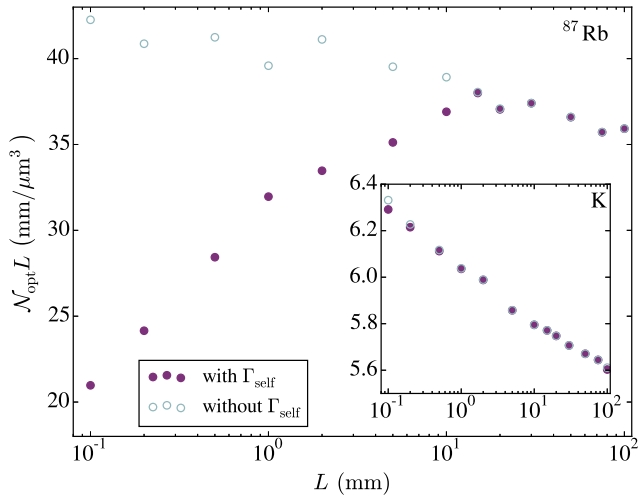}
\caption{Atomic number density after computer optimisation $(\mathcal{N}_\mrm{opt})$ multiplied by cell length $(L)$, as a function of $L$. The optimisation involves changing cell magnetic field and temperature of the cell in order to maximise the figure of merit at each cell length. The dark grey (purple) dots show the results when self-broadening is included in the model for the filter spectrum, while the light gray (blue) circles show the result without self-broadening. The main panel shows results for an isotopically pure $^{87}$Rb vapour while the inset gives the results for potassium at natural abundance.}
\label{fig:DopBufComp}
\end{figure}
The figure of merit of equation~\eqref{eq:FOM1} was maximized while simulating an isotopically pure $^{87}$Rb vapour with $L=100\,$mm, finding the optimal values of $B$ and $T$ to be $67.3\,$G and $60.9\,^\circ$C respectively. We then used the simple approach (section~\ref{sec:Simple}) to find the new values of the vapour cell temperature for a range of shorter cell lengths, and then evaluated the figure-of-merit values. In addition the figure-of-merit values were re-optimized (section~\ref{sec:CompOpt}) for each cell length to see if further improvement could be found. Figure~\ref{fig:FomOpt} shows the comparison of the two methods. We can see that the figure of merit changes with cell length, as is expected, since line broadening means that the filter spectra cannot be made identical for different cell lengths. We can also see that moving to shorter cells has a deleterious effect, but can be somewhat mitigated by re-optimization at each cell length.
\begin{figure}
\includegraphics[width=\linewidth]{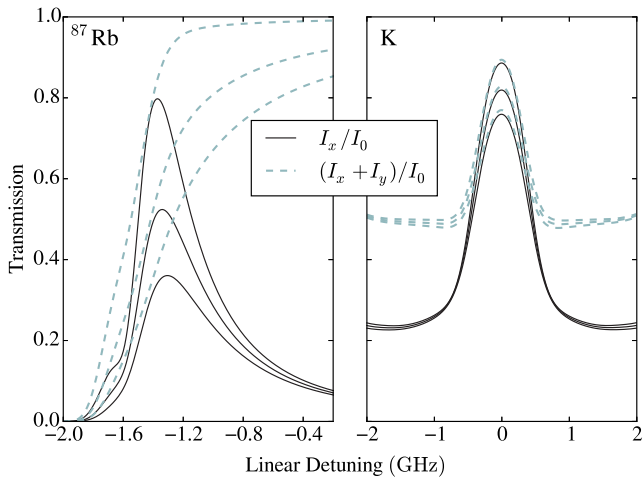}
\caption{Filter transmission ($I_x/I_0$, solid black curve) and cell transmission ($(I_x+I_y)/I_0$, dashed blue curve) as a function of linear detuning $(\Delta/2\pi)$, zoomed around the region of peak transmission. The left panel models a $^{87}$Rb vapour on the D$_2$ line, while the right panel models the K D$_2$ line; both of length 1 mm. The cell parameters were set to $B=85.8\,$G and $T=127.8\,^\circ$C ($\mathcal{N}=3.2\times 10^{13}\,\mrm{cm}^{-3}$) for $^{87}$Rb, and $B=864\,$G and $T=136.1\,^\circ$C ($\mathcal{N}=6.0\times 10^{12}\,\mrm{cm}^{-3}$) for K. The uppermost lines were calculated with a Lorentzian width given by natural broadening only ($\sim 6\,$MHz) while the middle and lower lines have a further 50 and 100 MHz of Lorentzian width. The global line-centres occur at 384.23042812~THz~\cite{Barwood1991,Ye1996} for the Rb D$_2$ line and 391.01617854~THz~\cite{Falke2006} for the K D$_2$ line.}
\label{fig:TransNfilter}
\end{figure}
\begin{figure}
\includegraphics[width=\linewidth]{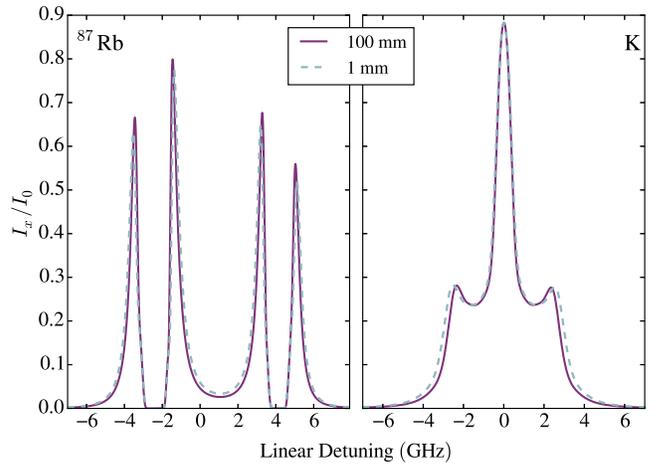}
\caption{Computer optimized Faraday filter spectra as a function of linear detuning. The optimal parameters were found to be $B=67.3\,$G and $T=60.9\,^\circ$C for the 100 mm long $^{87}$Rb vapour, $B=85.8\,$G and $T=127.8\,^\circ$C for the 1 mm long $^{87}$Rb vapour, $B=801\,$G and $T=70.2\,^\circ$C for the 100 mm long K vapour, and $B=864\,$G and $T=136.1\,^\circ$C for the 1 mm long K vapour. The ENBW is 2.0 and 2.2 GHz for the $^{87}$Rb vapour at 100 and 1 mm length receptively, whereas for the K vapour the ENBW is 2.4 and 2.6 GHz at 100 and 1 mm length receptively.}
\label{fig:WingVCentre}
\end{figure}

The inset of figure~\ref{fig:FomOpt} shows the result of a similar analysis for a potassium vapour at natural abundance~\cite{Rosman1998}, this time using the figure of merit of equation~\eqref{eq:FOM2} to produce a line-centre profile filter. The main difference in the results is that the figure of merit is less affected by decreasing cell length than the wing-type filter.

The reason for the difference between wing-type and line-centre filters can be elucidated by plotting the $\mathcal{N}L$ product as a function of $L$ after computerized optimization at each cell length, as shown in figure~\ref{fig:DopBufComp}. By repeating the optimization with the effect of self-broadening `turned off', we can see that the $^{87}$Rb wing-type filter is affected far more by self-broadening than the K line-centre filter. One can understand this difference in the behaviour of the two types of filters by inspection of the spectra (see figure~\ref{fig:TransNfilter}). Increases in Lorentzian broadening cause a decrease in transmission through the vapour cell at the filter frequency. This happens far more for the wing-type than line-centre filters.
\begin{figure*}
\includegraphics[width=0.8\linewidth]{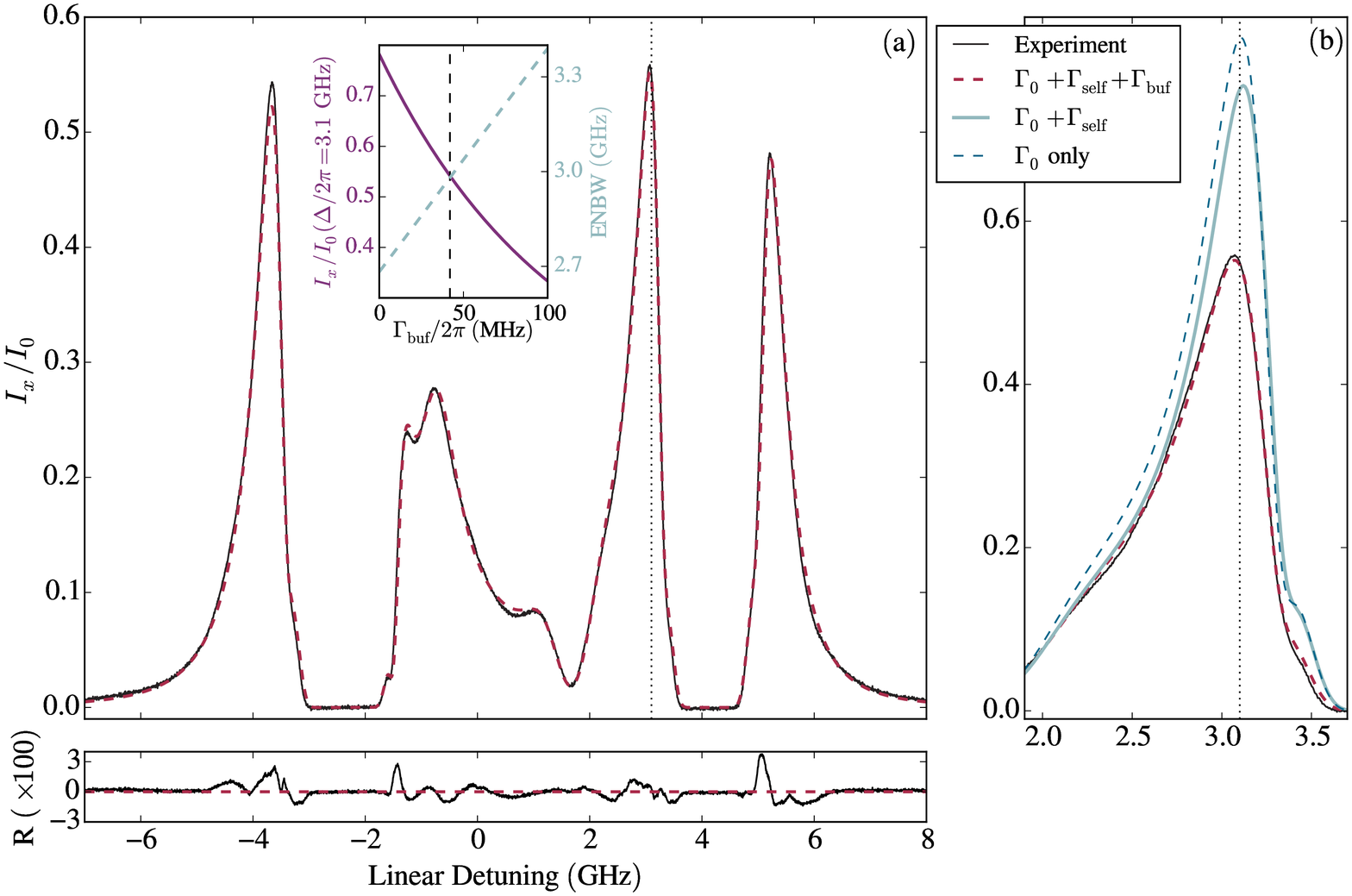}
\caption{Experimental and theoretical Faraday-filter spectra on the rubidium D$_2$ line as a function of linear detuning ($\Delta/2\pi$) from the weighted line-centre (384.23042812~THz~\cite{Barwood1991,Ye1996}). A 1 mm length vapour cell was used with an isotopic ratio of 99\% $^{87}$Rb to 1\% $^{85}$Rb. The solid black line in panel (a) shows the experimental filter spectrum and the dashed (red) line shows the fit to theory that includes the natural, self, and buffer gas induced ($\Gamma_\mrm{buf}$) Lorentzian broadening effects. Below panel (a) the residuals, R, (the difference between experiment and theory) are plotted. There is an RMS deviation between experiment and theory of 0.6\%. The inset of panel (a) shows the effect of $\Gamma_\mrm{buf}$ on transmission (solid purple line) and ENBW (dashed blue line) of theoretical filter spectra. The vertical dashed line marks the amount of buffer gas broadening seen in the experiment. Panel (b) shows a zoomed in region around the peak at 3.1 GHz, including theoretical curves with natural homogeneous broadening only (dashed blue) and with natural and self-broadening (solid blue).}
\label{fig:RbBuffer}
\end{figure*}
Changes in transmission on the wing of an absorption resonance due to Lorentzian broadening is due to the fact that Gaussian broadening decreases much faster than Lorentzian broadening with detuning from resonance~\cite{Siddons2009}. A higher optical depth transition feature will show this effect more strongly. This is one of the differences between wing and line-centre type filters. Wing-type filters rely on the sharp decrease in transmission caused by the atomic resonances to create narrow filter transparencies. This means that the circular dichroism cannot be too large since both polarizations need be scattered in the cell to sharply reduce the filter transmission to zero. However, a small amount of dichroism means that there is a small relative birefringence, which means that a high number density is required to create the large absolute birefringence necessary for the rotation of $\pi/2$. Conversely, the line-centre filter works by having a large circular dichroism, such that the transitions which absorb each polarisation of light are almost completely separated. We can see this in figure~\ref{fig:TransNfilter} where there the cell transmission is optically thick for just one circular polarization on either side of the transparency (causing $\approx50\%$ transmission of linearly polarized light through the cell and $\approx25\%$ transmission though the filter). This large dichroism comes with a large relative birefringence, meaning that the number density can be lower for a line-centre filter.

Line broadening clearly has a deleterious effect, however, good filter spectra for shorter vapour cells can be found so long as we change both the $B$ and $T$ to re-optimize the filter. This is shown in figure~\ref{fig:WingVCentre} where it is evident that the optimal filters achieved for a 1 mm cell length closely match that at 100 mm length.

\section{Experiment}\label{sec:Exp}

To compare theory with experiment for a compact cell, we used a micro-fabricated $1\times1\times1\,$mm$^3$ isotopically enriched $^{87}$Rb cell~\cite{Knappe2005}. The isotopic abundance of $^{85}$Rb was found by transmission spectroscopy to be $(1.00\pm0.02)\%$, in a similar way to that shown in ref.~\cite{Weller2012c}. This isotopic impurity affects the filter spectra, therefore the filter parameters were optimized taking this into account. We found the optimal parameters to be $B=72.0\,$G and $T=137.5\,^\circ$C, which gave a transmission peak at a detuning of 3.1 GHz.

The experimental Faraday filter arrangement is illustrated in figure~\ref{fig:setup}. The cell was placed in an oven to heat the cell near the optimal temperature, while the applied axial magnetic field was produced using a pair of permanent ring magnets. The field inhomogeneity across the cell was less than 1\%. Two crossed Glan-Taylor polarizers were placed around the cell to form the filter. A weak-probe~\cite{Smith2004,Sherlock2009} beam from an external cavity diode laser was focussed using a lens (not shown in figure~\ref{fig:setup}) with a 30 cm focal length, and was sent through the filter such that the focus was approximately at the location of the cell. After the filter, the beam was focussed using a 5 cm focal length lens onto an amplified photodiode. The laser frequency was scanned across the Rb D$_2$ transition, and was calibrated using the technique described in ref.~\cite{Siddons2008}.

Panel (a) of Figure~\ref{fig:RbBuffer} shows the experimental filter spectrum plotted with a fit to theory using ElecSus~\cite{Zentile2014a}. The fit parameters were found to be $B=73\,$G and $T=138.5\,^\circ$C. The first thing to note is that, due to the 1\% $^{85}$Rb impurity, the peak transmission occurs at $\Delta/2\pi=3.1\,$GHz rather than near -1.3 GHz if the cell were isotopically pure (see Figure~\ref{fig:WingVCentre}). Also, a further 42 MHz of Lorentzian broadening was added in addition to $\Gamma_0$ and $\Gamma_\mrm{self}$, due to the presence of a small quantity of background buffer gas in the vapour cell. This value was previously measured by transmission spectroscopy to be $\Gamma_\mrm{buf}/2\pi=(42\pm1)\,$MHz. Panel (b) of Figure~\ref{fig:RbBuffer} shows the filter spectrum zoomed into the main peak. In addition to the experimental and theory fit is the filter spectrum for the optimization that did not include the buffer gas broadening. We can see that the additional broadening drastically affects the filter transmission. Also by removing the effect of self-broadening from the theory, we again see a larger transmission. Table~\ref{tab:BroadeningComps} quantitatively compares the transmission, ENBW and FOM values for the curves shown in figure~\ref{fig:RbBuffer}. The inset of Panel (a) shows the filter transmission at a detuning of 3.1 GHz and the ENBW as a function of $\Gamma_\mrm{buf}$. The transmission decreases while the ENBW increases, showing that the performance (as measured by the ratio transmission to ENBW) of this kind of Faraday filter deteriorates quickly with increasing buffer gas pressures.
\begin{table}
\caption{Maximum transmission ($T_\mrm{max}$), equivalent-noise bandwidth (ENBW) and their ratio (FOM) for a 1 mm long isotopically enriched vapour cell. The magnetic field and temperature were 73 G and 138.5$^\circ$C respectively. The first row represents the fit to the experiment shown in figure~\ref{fig:RbBuffer}, while subsequent rows give the values after certain physical effects were removed (theoretically).}
\begin{center}
\begin{tabular}{cccc}
\hline
Spectrum & $T_\mrm{max}$ & ENBW (GHz) & FOM (GHz$^{-1}$) \\
\hline
Fit to Experiment \rule{0pt}{3.0ex} & 0.55 & 3.0 & 0.18 \\
No buffer gas \rule{0pt}{3.5ex} & 0.77 & 2.6 & 0.29 \\
\begin{tabular}{c} No self-broadening \rule{0pt}{3.5ex}\\ or buffer gas \end{tabular}& 0.83 & 2.6 & 0.31\\
\hline
\end{tabular}
\end{center}
\label{tab:BroadeningComps}
\end{table}

The amount of broadening due to buffer gas pressure that we observe, typically corresponds to approximately 1-2 Torr of buffer gas~\cite{Rotondaro1997,Zameroski2011}. The fact that this small pressure affects the filter spectra by a large amount shows that the wing-type Faraday filter spectra are very sensitive to buffer gas pressure. It has previously been shown that non-linear Faraday rotation can be a sensitive probe of buffer gas pressure~\cite{Novikova2002}, being non-invasive and using a simple apparatus. Our results show that it may be possible to use the linear Faraday effect instead, for which it is easier to model the effect of buffer pressure. However, it is not yet clear if this is more sensitive than using transmission spectroscopy~\cite{Wells2014}.

\section{Conclusions}\label{sec:Conc}

We have described an efficient computerized method to optimize the cell magnetic field and temperature for short cell length Faraday filters. From theoretical spectra we see that wing-type filters in particular are deleteriously affected by homogeneous broadening, while line-centre filters are less affected. We perform an experiment to realise a wing-type filter using a micro-fabricated 1 mm length $^{87}$Rb vapour cell, and find excellent agreement with theory. While buffer gasses can enhance some signals using vapour cells~\cite{Brandt1997}, they should be kept to a minimum in order to achieve the narrowest Faraday filters with the highest transmission.

\begin{acknowledgments}
We thank W. J. Hamlyn for his contribution to the experiment. We are grateful to S. Knappe for providing the vapour cell used in the experiment. We acknowledge financial support from EPSRC (grant EP/L023024/1) and Durham University. RSM was funded by a BP Summer Research Internship. The data presented in this paper are available from \url{http://dx.doi.org/10.15128/kk91fk598}.

\end{acknowledgments}

\bibliography{library}

\end{document}